\relax
\pdfoutput=1 
\documentclass[letterpaper]{article}
\usepackage{aaai18_modified}  
\usepackage{times}  
\usepackage{helvet}  
\usepackage{courier}  
\usepackage{url}  
\usepackage{graphicx}  
\begin{document}

\title{Effects of Social Bots in the Iran-Debate on Twitter}
\author{Andree Thieltges, Orestis Papakyriakopoulos, Juan Carlos Medina Serrano,  Simon Hegelich\\
Bavarian School of Public Policy\\
Technical University of Munich}
\maketitle
\begin{abstract}
2018 started with massive protests in Iran, bringing back the impressions of the so called ``Arab Spring" and it's revolutionary impact for the Maghreb states, Syria and Egypt. Many reports and scientific examinations considered  online social networks (OSN's) such as Twitter or Facebook  to play a critical role in the opinion making of people behind those protests. Beside that, there is also evidence for directed manipulation of opinion with the help of social bots and fake accounts. So, it is obvious to ask, if there is an attempt to manipulate the opinion-making process related to the Iranian protest in OSN by employing social bots, and how such manipulations will affect the discourse as a whole. Based on a sample of ca. 900,000 Tweets relating to the topic ``Iran" we show, that there are Twitter profiles, that have to be considered as social bot accounts. By using text mining methods, we show that these social bots are responsible for negative sentiment in the debate. Thereby, we would like to illustrate a detectable effect of social bots on political discussions on Twitter. 
\end{abstract}

\section{Bias in Political Discussions on Twitter}
By now, the existence of social bots in OSN's is indisputable \cite{Boshmaf2011,Boshmaf2013,Ferrara2016,Hegelich2016,Echeverria2017}. Moreover, there is plenty of software that allows an easy control of fake accounts in an automatized way. The ongoing use of such software is also documented by the continuous effort of the platform operators to monitor social bots accounts and their activity (e.g. Twitter audit). Political events discussed on Twitter, and the potential influence that social bots have on these discussions, is a fast growing research field, because the way people keep themselves informed about politics and cooperate to express their political opinion is changing. At the same time, these new ways and forms ``facilitate widespread exposure to political falsehood" \cite{Weeks2018}. Moreover, the ``presence" of social bots in political debates continuously grows since they have been identified as ``the main tools for online astroturf - or fake grassroots - smear[ing] campaigns during political moments world wide" \cite{Woolley2018}. 

Regarding to the topic ``Iran", a potential bias of political discussions and political activism on Twitter and other OSN's was discussed before: In 2009, the outcome of the Iranian presidential elections and the victory of Mahmoud Ahmadinejad caused nationwide unrests and protests. As these protests grew, the Iranian government shut down foreign news coverage and restricted the use of cell phones, text-messaging and  internet access. Nevertheless, Twitter and Facebook ``became vital tools to relay news and information on anti-government protest to the people inside and outside Iran"\footnote{https://www.counterpunch.org/2009/07/02/iran-networked-dissent/}. While Ayatollah Khameini characterized the influence of Twitter as ``deviant" and inappropriate on Iranian domestic affairs\footnote{https://www.wired.com/2009/06/tehran-threatens-bloggers-deviant-news-sites/}, most of the foreign news coverage hailed Twitter to be ``a new and influential medium for social movements and international politics" \cite{burns2009}. Beside that, there was already a critical view upon the influence of OSN's as ``tools" to express political opinion: Reports of ongoing attempts to manipulate the election or adding fuel to the flames of the following protests, cast a damning light upon the opportunities of political cyber-activism.
\begin{quote}
On the one hand, 'citizen reporters' found they could share stories with people around the world in a matter of minutes. On the other hand, 'trolls', 'vandals', 'rats', 'sock puppets', and other malicious online actors sought to spread false reports. The war in the streets spread to an online war of words. \cite{Carafano2009} 
\end{quote}
On the turn of the year 2017-2018, history seemed to have repeated itself as anti-government protests took place in more than 80 Iranian cities: Triggered by ``economic issues", the unrest started on December 30th and went on for almost a week until Iran's Revolutionary Guards intervened and ended up the ``sedition 1396"\footnote{http://edition.cnn.com/2018/01/03/asia/iran-protests-government-intl/index.html}. Regarding the political discussion on Twitter, Donald Trump's tweets accelerated the political debate in several hashtags by sympathizing with the anti-government activists, claiming to ``respect peoples rights" regarding freedom of expression and free internet access and by warning the Iranian government that ``the world is watching". Additionaly one of the tweets stated that it was ``time for a change"\footnote{https://www.theguardian.com/commentisfree/2018/jan/04/donald-trump-tweets-iran-protesters-sanctions} . In contrast to that, the Iranian leaders blamed the USA and Israel for meddling in its internal affairs:
\begin{quote}
Khamenei, speaking in the holy city of Qom, accused the United States and Israel of a ``carefully organized" plot to overthrow Iran's government. He said ``the enemy" had started chants against high prices, attracting Iranian demonstrators.\footnote{https://www.washingtonpost.com/world/irans-leader-blames-us-for-unrest-but-says-public-demands-must-be-answered/2018/01/09/7f1e9fde-f54b-11e7-beb6-c8d48830c54d\_story.html?utm\_term=.f81a0721ae6d} 
\end{quote}
Against this background of indications, given that    manipulation and bias is very likely in the political debate of the Iran protest, we analyzed approximately 900,000 tweets to detect social bots and deduce  statistical evidence from their behavior, proving an influential impact on  the specific political discussion. In our analysis we focused on the dissection of texts tweeted by social bots, in the context of the Iran protest.       
Still, there isn't so much profound or statistically firm knowledge on social bots as ``influencers" in political discussions on OSN's: Surveys in political science often aim at proving the existence or the overall quantity of social bots in a specific political discussion as statistical evidence of manipulation. The detection of social bots is per se a very complicated task, as they are designed to appear as accounts of human users \cite{Thieltges2016}. Therefore, research on social bots usually stops at their uncovering.  Assuming that the designed camouflage of social bots \cite{Abokhodair2015,Hegelich2015} is an insurance for a malicious influencer behind the machines, this decision might be right. But, since it is very unlikely to uncover the ``backers" of social bots, one can only speculate about a ``de-facto" influence or manipulation.

With our methodological analysis of the political discussion during the start of the protests in Iran, we'd like to rationalize the scientific discussions about social bots and their influence upon political debates on Twitter. Hence, we make the first step on quantifying the different behaviour social bots have in relation to human users.

\section{Related Work}
Analyzing social bots and their potential bias on political debates in OSN's has become an ever-growing research field in recent years. Currently, there is significant research on the impact of social bots relating to political events like elections: Kollanyi, Howard and Wolley \shortcite{Kollanyi2016} asserted, that ``political bots" are used to manipulate public opinion and that they are ``deployed" to polarize in ``sensitive political moments". Their analysis of Twitter hashtags based on the 2016 first U.S. Presidential Debate showed the presence of political bots in these discussions. Moreover, they found that human users on Twitter used ``more politically neutral" hashtags than social bots. 

Ferrara et. al \shortcite{Ferrara2016} analyzed the 2016 Presidential race in the USA during three periods, between 16 September and 21 October 2016, and concluded, that social bots distorted the online discussion on Twitter. Their findings suggested, 
\begin{quote}
that the presence of social media bots can indeed negatively affect democratic political discussion rather than improving it, which in turn can potentially alter public opinion and endanger the integrity of the Presidential election.
\end{quote}

Shao et al. \shortcite{Shao2017} researched the spread of misinformation in the context of the US-Presidential elections on Twitter. They found that it is very likely that social bots are “effectively boosting” the viral diffusion, by “actively amplifying” claims ``in the first few seconds after they are first posted". 

Corresponding to that, the analysis of disinformation and social bot operations in the run up to the 2017 French Presidential election advanced the hypothesis, ``that a black market of reusable political disinformation bots may exist". Ferrara \shortcite{Ferrara2017} identified the presence of social bots that already were detected in the US-Presidential elections. They actually were used ``to support alt-right narratives", then were shut down ``and came back into use in the run up days to the 2017 French presidential election".

Arnaudo \shortcite{Arnaudo2017} studied computational propaganda spread in Brazil by researching the 2014 presidential elections, the impeachment of former president Dilma Rousseff and the 2016 municipal elections in Rio de Janeiro. On basis of Twitter hashtags related to these events and their context, he analyzed how social bots operated in the discussions. He found that in the debates around the elections, as well as in the impeachment discussions, social bots were deployed by the opponents of the former president Dilma Rousseff.

Forelle et al. \shortcite{Forelle2015} also focused on researching social bots that have been used to shape the public opinion or spread political propaganda. By analyzing the users that retweeted the politicians content, they found that there are social bots that frequently pretended to be official political candidates. According to Arnaudo's study, the social bots mostly were used by a (radical) opposition party (VP). But in contrast to the social bots found in the Brazilian context, social bots in this study were mostly ``promoting innocuous political events" rather ``than attacking opponents or spreading misinformation". However, there is also evidence that social bots have been used to oppress and disrupt political debates on Twitter that are critical of the government: Surez-Serrato et al. \shortcite{Suarez2016} showed in their study on the \#YaMeCanse – hashtag\footnote{\#YaMeCance was the most active protest hashtag in the history of Twitter in Mexico.}, ``that bots played a critical role in disrupting online communication about the protest movement". In contrast, there is also research that doubts the influential effects of social bots on political events and the related discussions:  Murthy et al. \shortcite{Murthy2016} analyzed the interaction of humans and social bots in the discussions of the UK-general elections in 2015. Their experimental approach, including the use and observation of social bots, suggested that bots have “very little effect on the conversation network at all”. Nevertheless, Murthy et al. revealed that ``there are economic, social, and temporal factors that impact how a user of bots can influence political conversations". Graham and Ackland \shortcite{Graham2016} argued the malicious intention of social bots in political discussions: They developed a ``normative role" for social bots, which suggested that they are able to build ``bridges between separate, ideologically homogeneous subnetworks", expose ``tightly knit clusters of users to alternative viewpoints", or bring ``about measurable shifts towards deliberative democracy in online discourse." Taking the above in consideration, we analyze the effect of social bots in the Iran protests.

\section{Data}
The data used were generated via the Twitter streaming-API. We accessed this API for a period of 24 hours from 30th to 31th December 2017 and searched for the word ``Iran" everywhere in Twitter. Thereby, we aggregated 899,745 tweets and their metadata.  

\section{Detection of Social Bots}
Initially, some problems with the definition of social bots should be illustrated: The term ``social bot" indicates an automated account in an OSN that simulates a human user to camouflage its robotic nature. Usually, social bots are defined and categorized according to their intentions, as well as the way they imitate human behaviour \cite{Stieglitz2017}.  Another aspect that remains unclear is what level of automation an account needs to exhibit in order to be defined as social bot. Relating to the content that is posted on OSN's, it is quite hard to make a clear judgement, because human users may use apps like Tweetdeck etc. to spread their ``handcrafted" content automatically (e.g. via Twitter API). At the same time, one does not need any software to achieve a high-level automation, e.g. the case that  human user groups generate a mass of posts or tweets in high frequency by employing copy \& paste. Besides that, the fraudulent intent could only be assumed, because the specific intention of the users or the ``backers" is unknown. Thus, it is almost impossible to make an undoubted social bot detection. 

In research activities related to social bots two different detection methods have been developed: Machine learning and heuristics. In general, the detection with machine learning methods is based on hand-coded data sets, which are automatically scanned for patterns. Hence, social bot classification predicated on machine learning methods often has to face the problem of similarity: Such systems can only detect social bots that are similar to the ones on which the classifier was trained.

On the other hand, heuristic approaches deal with theoretical deduced detection rules, that separates humans from bot. The drawback of this method is the rigidity of the rules: Systems that detect social bots by employing heuristic rules often lack an improvement. Moreover, it is very hard to identify rules that will work for the detection of specific social bots. In addition, the combination of different rules leads to interdependencies that may affect the detection process in general. 
In the case of our data-set of tweets and metadata we chose a heuristic approach for the detection of social bots. This methodological decision was based on two crucial facts:
1. It is not clear that human coding is a gold standard for training data. The better the camouflage of the social bots, the harder it will be for humans to differentiate between bot and human user. There are approaches reporting the inter-coder-reliability \cite{Varol2017}, but these approaches only tell us how likely it is that different humans came up with the same classification. This does not rule out that there is a class of bots that is systematically overseen by human coders. 
2. The successful machine learning approaches all include many variables that are taken from the meta-data of the tweets and thereby are invisible for human coders. 

For our detection analysis we used four different heuristics:

\subsection{1. Analysis of the Source}
There is a note in the metadata of every tweet, whether it has been sent via the Twitter app or any other app that connects via the Twitter API. Our analysis of this source variable unveiled 97 ``suspicious" app services that connects via Twitter API within the Iran data set. Some of these apps are well-known services for the automated access to Twitter, used also by respectable and reliable accounts (e.g. IFTTT). To some extent, obvious social bot software (e.g. twittbot) or rather dubious sources were found (www.AgendaofEvil.com, pipes.cyberguerril.org or www.rightstreem.com). There are also alternative Twitter apps and media channels found as sources. It is uncertain though, if these source can be used for running social bots. Overall, 24,718 Tweets have been sent using such sources.

\begin{figure*}
  \includegraphics[width=\textwidth,scale=0.2]{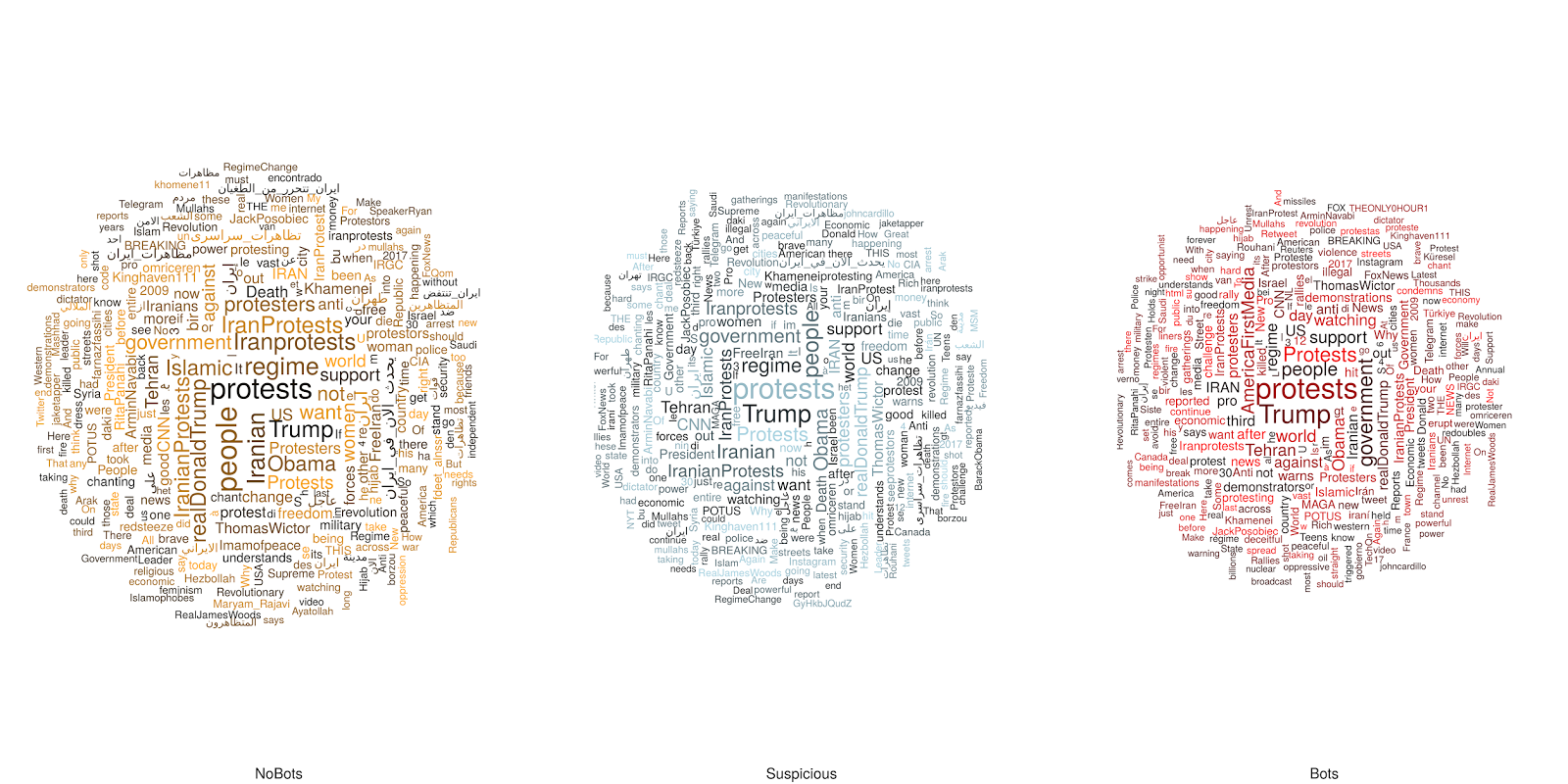}
  \caption{Word clouds from the three categories of classification for tweets}
\end{figure*}

\subsection{2. Friend-Follower-Ratio}
The friend-follower-ration is one of the most interesting heuristic relating to Twitter: In the past social bots used to have an above average friends rate and very few followers of their own \cite{Wang2010}. The idea of most of the bot-herders was to have as many friends as possible and therefore spread the bot content as fast and as wide as possible. With this in mind, Twitter started monitoring the so called ``aggressive following and follow churn (repeatedly following and unfollowing large numbers of other accounts)". Every account on Twitter is now limited to follow 5000 accounts in total. There is the possibility of exceeding this limit, by fulfilling an extra hidden criterion: Twitter checks the followers to following ratio, and decides if someone can follow additional accounts.\footnote{ https://help.twitter.com/en/rules-and-policies/twitter-following-rules.} Our research within the field of social bot detection shows that most social bot accounts on Twitter nowadays overcame this limitation rule by having an almost equalized friend-follower-ratio. They manage to do so, because they connect (make friends and followers vice versa) themselves with other bots. Nevertheless, there are human users on Twitter who also have an equalized friend-follower-ratio by chance. Still, most of the human Twitter-accounts had an average of approximately 100 followers, so we expanded our heuristic rule and just included those accounts with more than 100 followers and an equalized friend-follower-ratio. Within this method we identified 58,372 ``suspicious" tweets.

\subsection{3. Number of Tweets per Day}
The number of tweets per day posted by a single Twitter-account is another efficient criterion to classify humans and social bot-accounts on Twitter: It is a widely held belief that social bots have a very high activity rate in comparison to the average human user. But there is no clear definition of a conspicuous tweet-frequency: One could simply set a specific peak-value: E.g. Howard and Kollanyi \cite{Howard2016} use a peak-value of 50 tweets per day to identify social bots. In general though, this procedure tends to be arbitrary. To get a heuristic rule that is more accurate, one needs to calculate a critical value. There are two different approaches, that are suitable: 

1. Calculate the interquartile-range (IQR aka. the interval between the upper quarter and the median) and multiply the value by 1.5. This rate is a typical outlier-rate in analytical statistics \cite{Friedman2002}. All values that are one and a half times bigger than the IQR have a disproportional range relative to the median. Regarding our sample this includes all accounts that tweet more than 54 times a day.

2. A more conservative approach is to take the most active 5 \% users as conspicuous. We used this approach for our sample and calculated a peak-value of 165 tweets per day as a conspicuous tweet rate. We preferred this approach to minimize the error rate.

\subsection{4. Text Duplicates}
If text is automatically generated or spread in large quantities, there is a good chance that identical texts appear. If one shares or cites a piece of text on Twitter it is usually marked as a so called ``retweet". If identical texts appear without being retweeted, this may be an evidence for automated text duplication. According to our sample the number of identical or duplicated texts were relatively low (820 tweets). 

\subsection{Combination of Heuristics}\label{sssec:num1}
When different heuristic rules are used, there is always the question of the ratio between the single rules: Is it mandatory that all rules must be fulfilled all the time, or is it enough, if any rule is fulfilled? As said before, social bots are very different to each other and one could lose detection accuracy, if only the cases which fulfill all the rules are analyzed. On the other hand, if any of the rules is suitable to identify a social bot, one has to deal with the rule-specific error rate. E.g. if we take our ``number of tweets per day" rule at face value, we have to assume a priori that 5\% of the most active users are social bots. Against the background of these ``accuracy trade-offs", we chose to work with a three-step procedure: First we classified all of the tweets that apply to any of our rules as ``suspicious". Second, we took all tweets where at least two different rules applied and labeled these tweets as ``social bots". In both sets, we sorted out all of the tweets made by ``verified accounts", which are often media accounts or news channels. Media accounts on Twitter could be easily misclassified as bots, because of their high-frequency tweets and the fact that most of them are automatized. Usually these accounts are verified by Twitter. In a third step these verified accounts are excluded from the social bot classification. By combining the heuristics, we finally got three categories of classification for our tweets: No bots, Suspicious and Bots.

\section{Text Mining Methods}  
\begin{figure}
	\centering
	\includegraphics[scale=0.18]{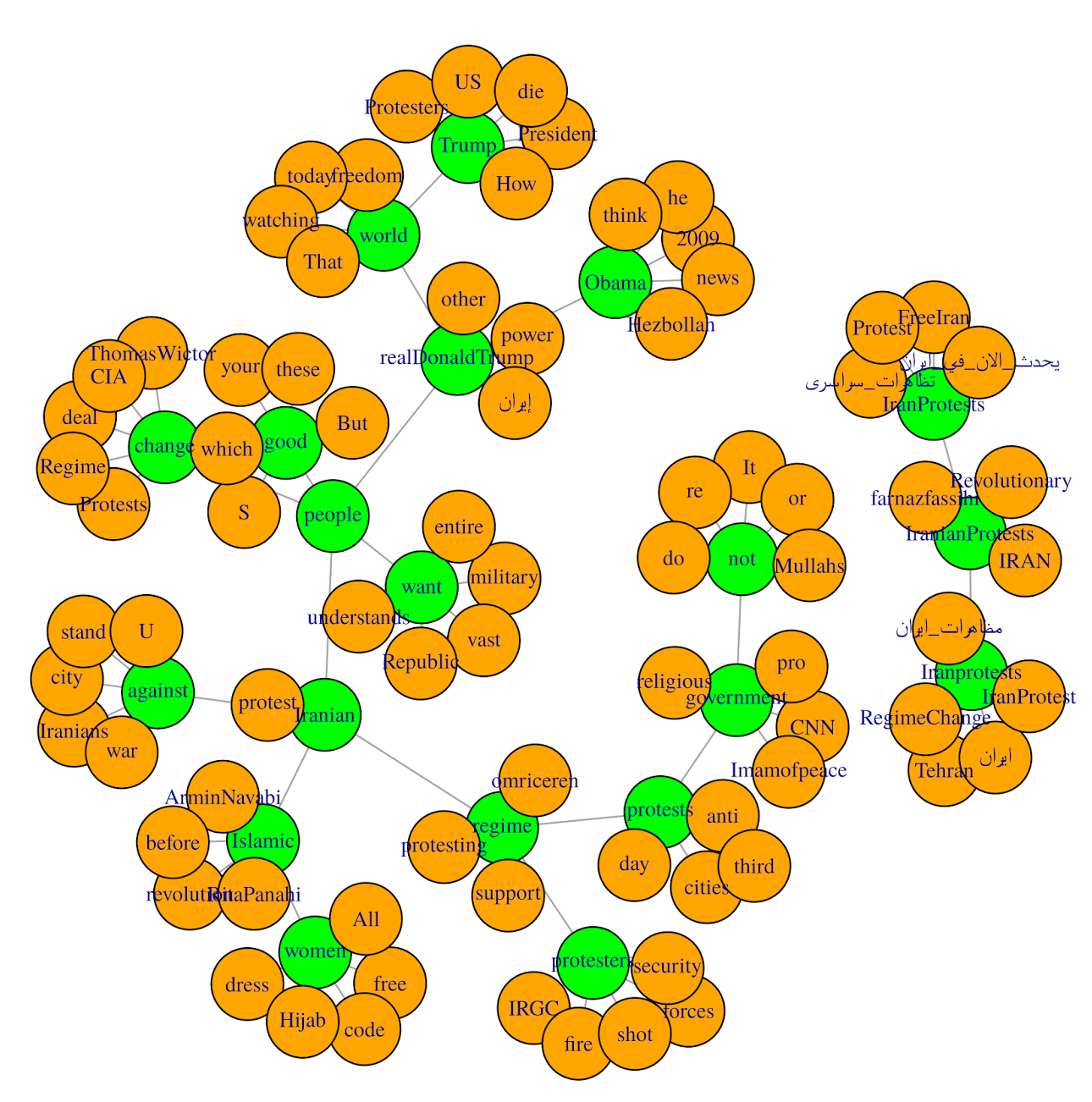}
	\caption{Co-ocurrance graph for tweets from normal users}
\end{figure}

\begin{figure}
	\centering
	\includegraphics[scale=0.18]{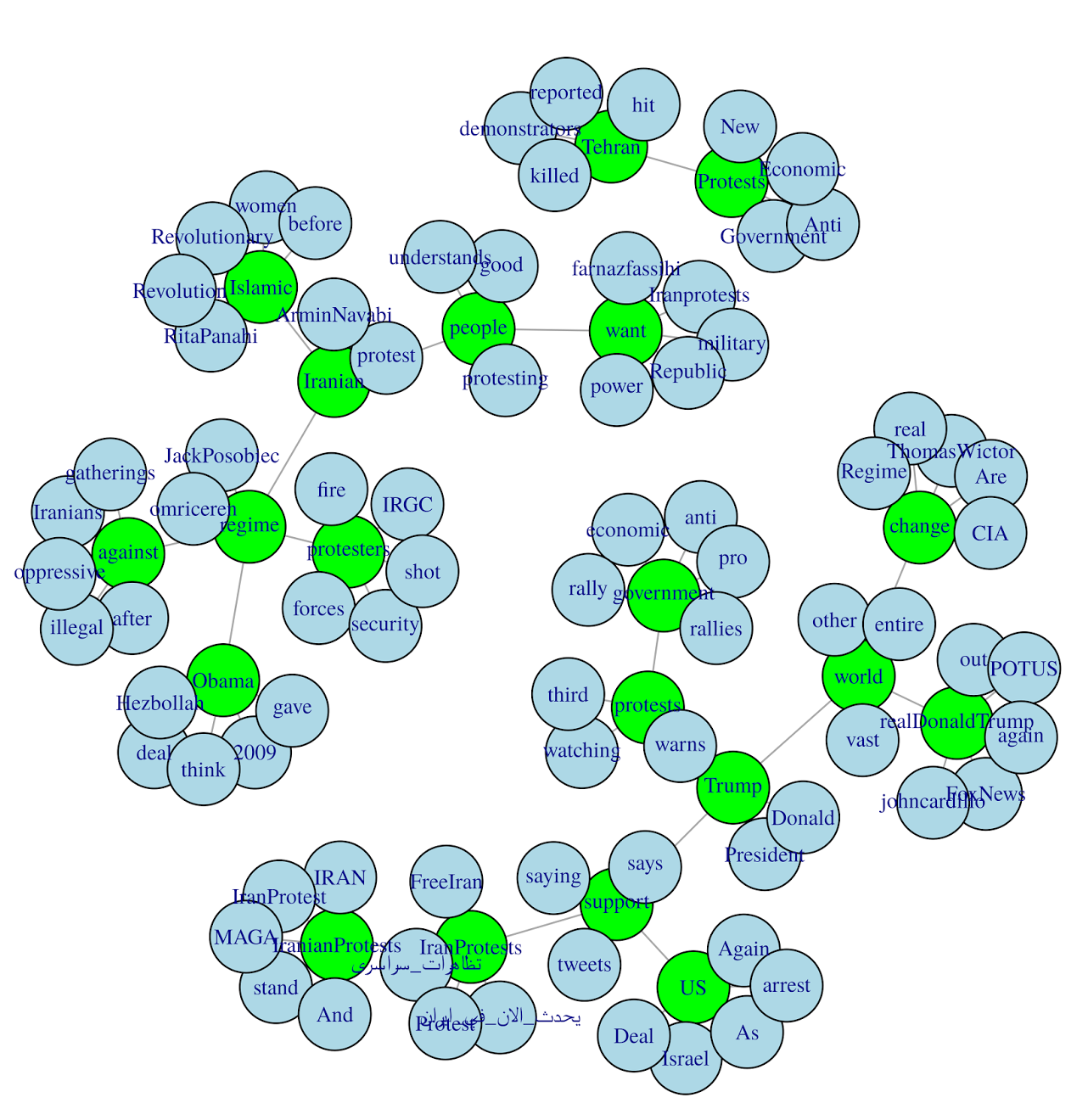}
	\caption{Co-ocurrance graph for tweets from accounts identified as suspicious}
\end{figure}

In the first step we employed a ``bag-of-words"-approach to analyze the text of the tweets in our sample: It counts the frequency of single words in every single tweet and displays the result as a word-cloud: The bigger a word is displayed in the word-cloud, the higher its frequency. Figure 1 shows the three different word-clouds for our three different categories of tweets (No Bots, Suspicious, Bots). A high frequency for a specific word in a huge mass of different documents doesn't mean that this word is very significant to a specific document. On the contrary, single words that could be found very frequently in one specific document are very often significant for this document. To avoid this problem of simple frequency distributions we employed the ``term frequency-inverse document frequency" (TFIDF) distribution on our sample \cite{Aizawa2003}. Before employing our ''bag-of-words'' approach, we removed standard stop-words as English articles as well as the term "Iran". Furthermore, we pruned the vocabulary of words, by taking into consideration only words that appear at least in  1\% and at most in 45\% of the documents. Pruning the vocabulary has a significant impact in the accurate context recognition of documents \cite{Madsen2004}.

\begin{figure}
	\centering
	\includegraphics[scale=0.18]{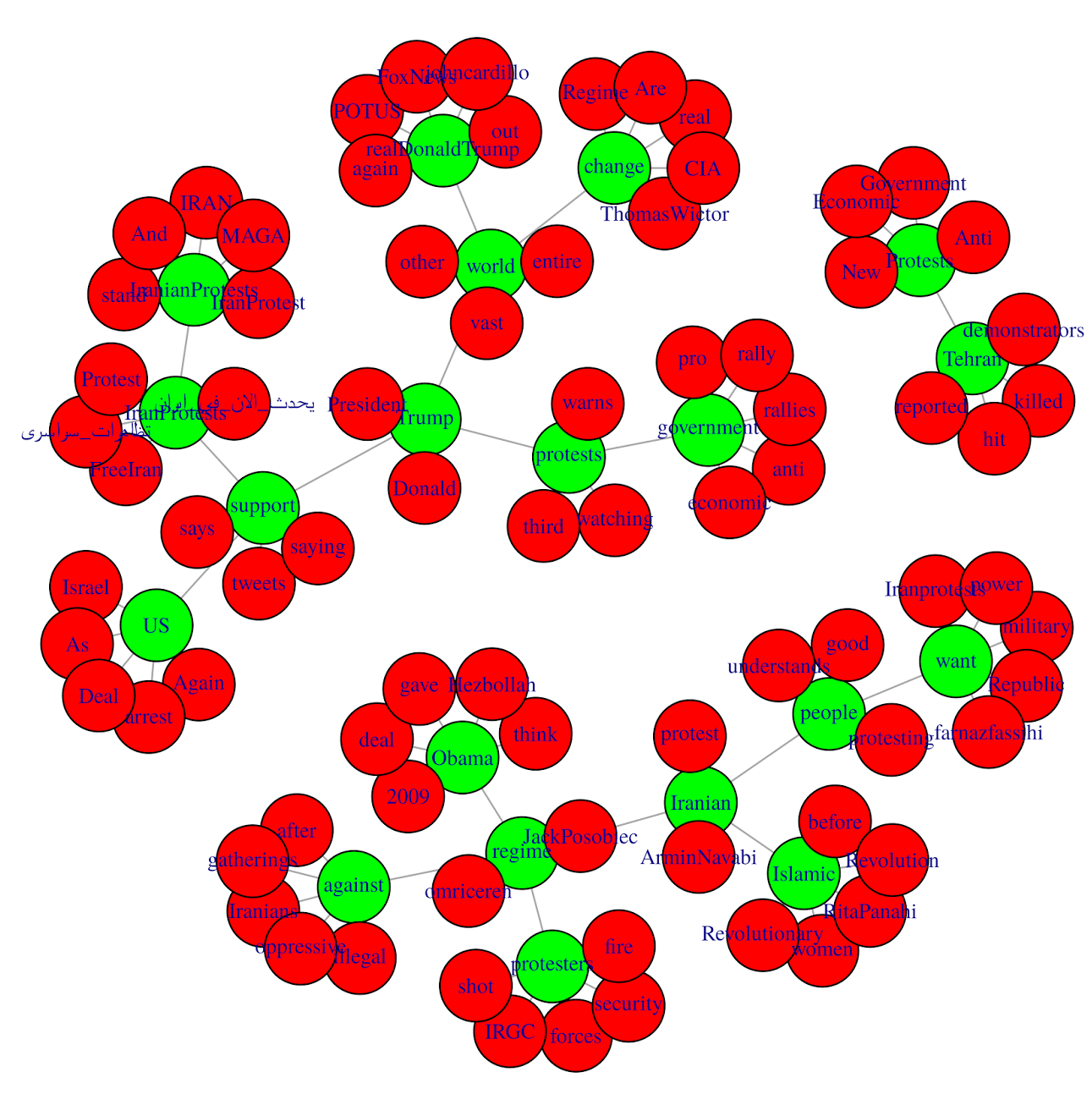}
	\caption{Co-ocurrance graph for tweets from accounts identified as bots}
\end{figure}

\begin{table*}
  \begin{center}
    
    \label{tab:table1}
    \begin{tabular}{l|r|r|r|r} 
      \textbf{Screen Name} & \textbf{Tweets per Day} & \textbf{Follower} & \textbf{Friends} &\textbf{Source}\\
      \hline
        Davewellwisher &	1082 &	27374	& 15854 & IFTTT\\
        TinaOrt79591465 &	291	& 7492 &	7841 & Twitter for Android\\
        americanshomer &	310 &	5636 &	5669 &	Twitter for iPhone\\
        BetigulCeylan &	231 &	2394 &	2277 &	Twitter for iPhone\\
        zyiteblog &	633 &	1688 &	3687 &	Hootsuite\\
        ErengwaM &	269 &	1177 &	1200 &	Twitter for Android\\
        PeggyRuppe	 & 230 &	6424 &	6347 &	Twitter for iPhone\\
        sturm\_tracey &	234 &	2058 &	2105 &	Twitter Lite\\
        CityofInvestmnt &	97  &	5180 &	5017 &	IFTTT\\
        emet\_news\_press &	519 &	14137 &	3304 &	Hootsuite\\
        favoriteauntssi &	476 &	5045 &	5166 &	Twitter for iPhone\\
        Sakpol\_SE &	442 &	4605 &	0 &	Sakpol Magic\\
        dreamedofdust &	626 &	2640  &	84 &	IFTTT\\
        NarrendraM &	2034 &	1952 &	251	 & IFTTT\\
        YMcglaun &	207 &	10571 &	10559 &	Twitter for Android\\
        lynn\_weiser &	406 &	20009 &	19756 &	Twitter for iPhone\\
        AngelaKorras &	194 &	3666 &	3768 &	Echofon\\
        MarjanFa1 &	262 &	6183 &	6059 &	Twitter for Android\\
        RichGossger &	355 &	587 &	571 &	Twitter Web Client\\
        sness5561\_ness &	451 &	11994 &	12477 &	Twitter for iPhone\\
    \end{tabular}
    \caption{Top 20 of the most active accounts that were classified as bots. The column \textit{tweets per day} is rounded to the nearest integer and the \textit{source} column represents the source extracted from the url}
  \end{center}
\end{table*}

In a second step, we analyzed the words that frequently ``co-occure" with each other to determine, if the same words optionally were used in different contexts. We created a term-co-occurrence matrix calculating the likelihood that two words appear in the same context. The context was defined as a skipping window of 5 words. For our three different categories of tweets we sort out the 20 most frequent words and displayed the 5 words that were most commonly used in the same context. The results of this procedure are displayed in three different figures (see Figure 2, 3, 4).  

In a third step we employed sentiment analysis on all the tweets in our three categories: All words in every single tweet were compared with a dictionary\footnote{https://cran.r-project.org/web/packages/qdap/index.html}, that identifies if every single word is rather used in a positive or negative way. Corresponding to the findings of the dictionary, every word is assigned to a positive or negative value. Adding up all the values from the words that constitutes a tweet, shows if the tweet has rather a positive or negative sentiment. Although more complex sentiment analysis methods exist, we preferred this simple straightforward one, as it’s efficiency in Twitter is proven \cite{Kouloumpis2011}. Due to the fact that an English dictionary was applied, the sentiment analysis does not consider any other languages. 
Finally, we analyzed if the different outcomes of our sentiment analysis were random or not: By reviewing the sentiment of all words in our different categories with the Kolgomorov-Smirnov-Test (KS-test): The null hypothesis is here, that all of the data is part of the same distribution function.

\section{Findings}  
From the total number of tweets, we identified 781,674 tweets  that came from human users (86.87\%), 118,071 from ``Suspicious"-tweets (13.12\%) and 10,126 from ``Bot"-tweets (1,12\%). The "Bot" tweets percentage is included in the "Suspicious" tweets, according to the methdology explained in the \textit{Combination of Heuristics} subsection.

Table 1 gives a clearer picture of the kinds of accounts that were identified. It shows the 20 most active users in the collection of tweets that were classified as Bots. The \textit{tweet per day} illustrates that they are very active accounts that differ from the normal users. Moreover, for most of the accounts, the number of followers and friends is very similar. From the 20 accounts, only the user \textit{Sakpol\_SE} identifies itself as a Bot in its profile description.

On the text mining side, we observed that the bag-of-words analysis and the word clouds for the ``No Bots", the ``Supicious" and the ``Bots" classified tweets didn't come up with wide differences: Indeed, the word ``Trump" seemed to be more frequent in the tweets of the ``Bots"-group. Furthermore, in the tweets generated by humans we can identify a higher use of persian vocabulary in the word clouds. Apart from these cases, the used words and their frequency in our three categories of tweets showed only narrow differences. Most of the words are related to the Iran protests, as well as U.S. Politicians, CIA, and news media.

The results of ``co-occurence"-test showed some differences between the three categories of tweets and a significant amount of similarity as well: As shown in all figures (see Figure 2, 3 and 4) there is a distinct connection between ``Obama" and ``Hezbollah", so the topic of blaming Obama was present among real users as well as among bots. The results of the ``Bots"-group shows that there is a tendency for economic topics  and that the ``Trump"-topic has a much higher significance compared with the ``No Bots" and the ``Suspicious"-groups. The hashtag \#MAGA (Make America Great Again) also appears in the ``Bots"-tweets. In contrast to the ``Suspicious" and the ``Bots"-tweets, the topic of ``Women's Rights" only occurred in the ``No Bots"-tweets. Overall, our result of the ``co-occurence"-analysis shows, that social bots and human users seem to tweet to very similar topics. Therefore, the contextual representation of the protests doesn't change that much. 

But there are significant differences between the tweets of the social bots and the human users: The result of the sentiment analysis showed, that the social bots used more negative words in their tweets. The average sentiment of the ``Bots"-group is -0.094. Compared to the ``No Bots"-group (-0.049) and the ``Supicious"-group (-0.063) the negative sentiment in every single social bot tweet is way higher than in the tweets of the other groups. Moreover, this result is not random: The KS-test showed a value of 0.12 and a p-value of 0.0, which reveals that it is very unlikely that the words (and their positive and negative sentiment values) are derived from the same distribution. By looking at the empirical cumulative density function (Figure 5), one can visualize these differences: The density function of the social bots has a distinct characteristic in the negative range.

\begin{figure}
	\centering
  \includegraphics[scale=0.15]{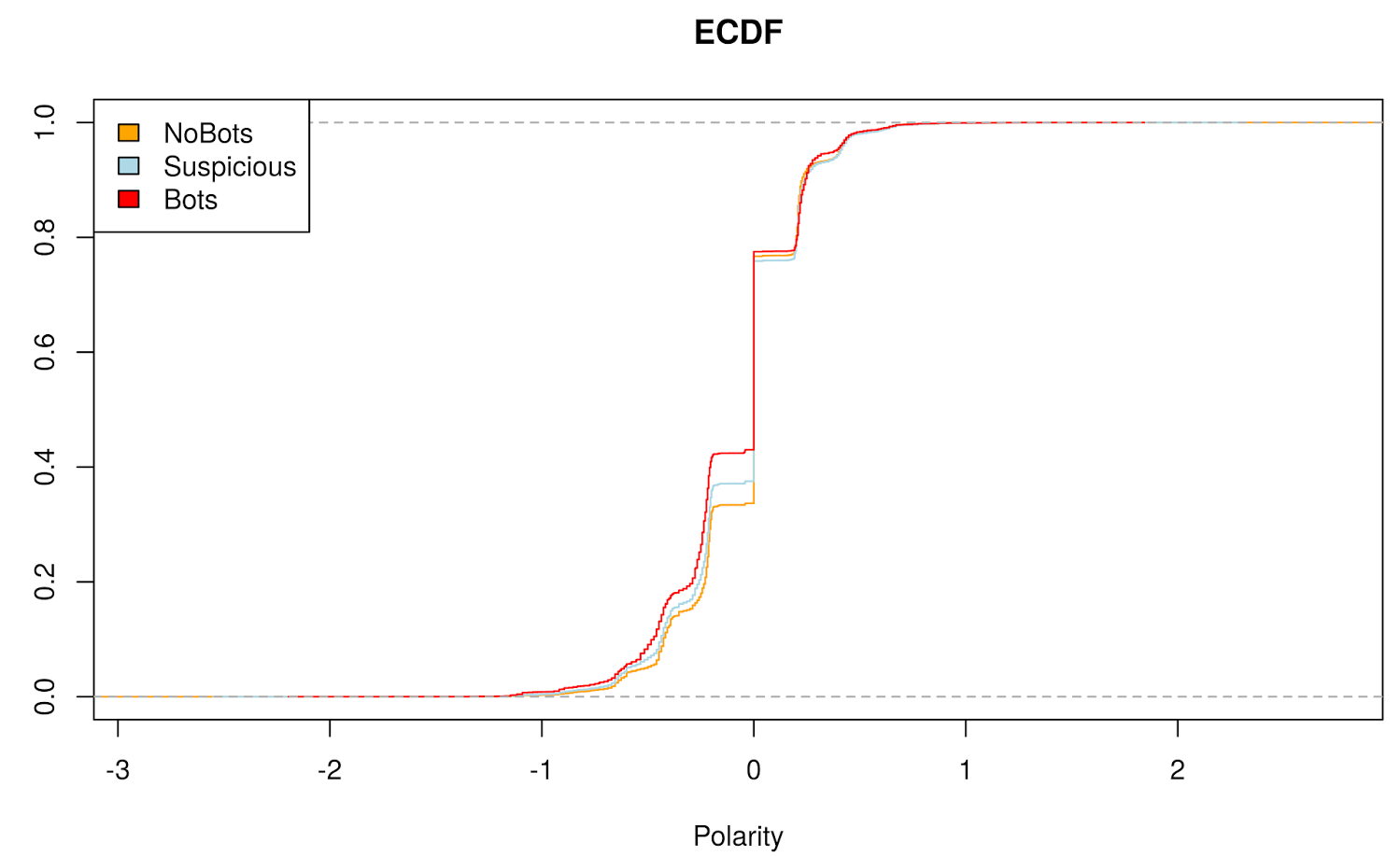}
  \caption{Empirical cumulative density function}
\end{figure}

\section{Conclusion}  
Social bots have a detectable impact on the atmosphere of political discussions measured as sentiment. If one follows the thesis that debates or discussions on Twitter have a political effect, and that the atmosphere within social networks becomes more and more important, then one has to conclude, that in the case of Iran protests, social bots have a political impact. This means, the existence of a measurable difference between the behavior of bots and humans is a necessary condition for any ``bot-effect": If the bots were just mirroring the  ongoing debate without any distortion, it would not be possible to differentiate if humans read the messages from bots or from other users. 

Now, the next step would be to find out, if humans who read the negative bot-texts change their behavior over time (and become more negative themselves or are more likely to leave the debate). Prerequisite for that would be a more extensive sentiment and text analysis of the tweets, in order to reveal additional differences of bot-generated tweets and their context. For example, topic modeling algorithms could be applied to detect content differences. Moreover, a subjectivity/objectivity analysis could reveal additional features that could lead to the influence of the public.  But this examination would be beyond the scope of the data used in this research.

Finally, it should be kept in mind that text mining as a research tool deals with a lot of implicit methodological rules -e.g. especially in case of the cleaning operations - that raised some concern about their affection on our results. 

\bibliographystyle{aaai}
\bibliography{bibliography.bib}

\end{document}